\documentclass[aps,pra,twocolumn,longbibliography, superscriptaddress,nofootinbib,10pt]{revtex4-1}
\usepackage[]{amsmath}
\usepackage{graphics}
\usepackage{amssymb}
\usepackage{graphicx,epstopdf}
\usepackage[usenames,dvipsnames,svgnames]{xcolor}
\usepackage{comment}

\AtBeginDocument{%
    \newwrite\bibnotes
    \def\bibnotesext{Notes.bib}
    \immediate\openout\bibnotes=\jobname\bibnotesext
    \immediate\write\bibnotes{@CONTROL{REVTEX41Control}}
    \immediate\write\bibnotes{@CONTROL{%
    apsrev41Control,author="08",editor="1",pages="1",title="0",year="1"}}
     \if@filesw
     \immediate\write\@auxout{\string\citation{apsrev41Control}}%
    \fi
}%

\begin{document}

\title{Do Gedankenexperiments compel quantization of gravity?}
\date{\today}
\author{Erik Rydving}
\email{erydving@kth.se}
\affiliation{School of Engineering Sciences, KTH-Royal Institute of Technology, SE-114 28 Stockholm, Sweden}
\author{Erik Aurell}
\email{eaurell@kth.se}
\affiliation{Department of Computational Science and Technology,
KTH-Royal Institute of Technology, AlbaNova University Center, 
SE-106 91 Stockholm, Sweden}
\author{Igor Pikovski}
\email{igor.pikovski@fysik.su.se}
\affiliation{Department of Physics, Stevens Institute of Technology, Castle Point on the Hudson, Hoboken, NJ 07030, USA }
\affiliation{Department of Physics, Stockholm University, AlbaNova University Center, SE-106 91 Stockholm, Sweden\vspace{0.2cm}}

\begin{abstract}
    Whether gravity is quantized remains an open question. To shed light on this problem, various Gedankenexperiments have been proposed. One popular example is an interference experiment with a massive system that interacts gravitationally with another distant system, where an apparent paradox arises: even for space-like separation the outcome of the interference experiment depends on actions on the distant system, leading to a violation of either complementarity or no-signalling. A recent resolution shows that the paradox is avoided when quantizing gravitational radiation and including quantum fluctuations of the gravitational field. Here we show that the paradox in question can also be resolved without considering gravitational radiation, relying only on the Planck length as a limit on spatial resolution. Therefore, in contrast to conclusions previously drawn, we find that the necessity for a quantum field theory of gravity does not follow from so far considered Gedankenexperiments of this type. In addition, we point out that in the common realization of the setup the effects are governed by the mass octopole rather than the quadrupole. Our results highlight that no Gedankenexperiment to date compels a quantum field theory of gravity, in contrast to the electromagnetic case. 
\end{abstract}

\maketitle

\section{Introduction}
A full quantum theory of gravity is a major outstanding challenge in modern physics. 
While much important work is focused on solving this challenge, one may ask whether quantizing gravity is logically necessary for consistency with known physics.
Prior to the development of quantum electrodynamics, the analogous problem of electromagnetic field quantization was considered by Landau and Peierls \cite{landau1931erweiterung}, and then famously by Bohr and Rosenfeld \cite{bohr1933frage}.
The latter work showed that electromagnetism requires field quantization to be consistent with the uncertainty principle applied to charged particles. This insight, based on a series of Gedankenexperiments, preceded the development of the full theory. A natural question is whether these arguments, or similar Gedankenexperiments, can be provided for a quantum field theory of gravity. It was realized early on by Bronstein that the Bohr-Rosenfeld argument does not apply to gravity \cite{bronstein1936quantentheorie}. He isolated two crucial differences that prevent a gravitational version of the Bohr-Rosenfeld setups: a single charge of gravity instead of two opposing charges, and the non-linear backaction of the probe system onto the measured gravitational field. Since then, arguments in favor of field quantization \cite{bergmann1957summary, eppley1977necessity} and against \cite{penrose1996gravity, dyson2014graviton} have been put forward. 
\\

\section{The Gedankenexperiment}
\begin{figure}[t]
\centering
\scalebox{0.43}{\includegraphics[]{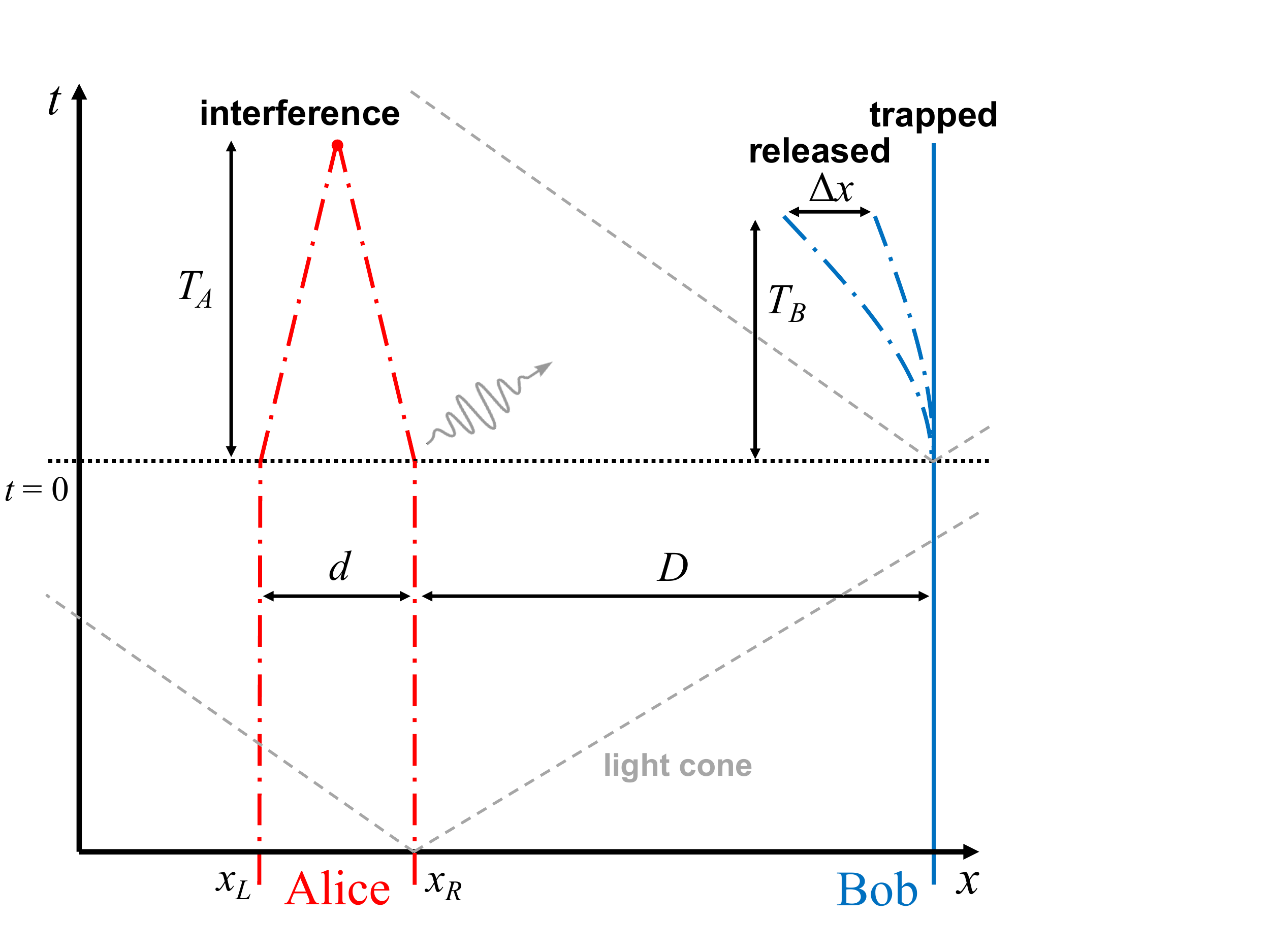}}
\caption[]{\textit{The Gedankenexperiment of interest (see main text for details). Alice performs an interference experiment with a massive system in a spatial superposition of separation $d$, while Bob at a distance $D$ chooses whether to release his particle from a trap or not. If released, the particle will experience a superposition of two gravitational accelerations towards Alice (since Alice prepared her superposition in the causal past), and will reach a final separation $\Delta x$ between the two states after a time $T_B$. 
When Alice's system is recombined in a time $T_A$ to get interference, it emits gravitational radiation. Independently, Alice requires a high enough precision to resolve her interference pattern.
}}
\label{fig:setup}
\end{figure}

In a recent paper, Belenchia et al. \cite{Belenchia2018} showed that a quantum treatment of the gravitational field solves a paradoxical situation in a Gedankenexperiment first introduced by Mari et al. \cite{mari2016experiments}. It is a variation of an argument first proposed by Niels Bohr, as outlined in detail by Baym and Ozawa \cite{Baym2009}. A similar argument is briefly stated in Feynman's lectures on gravitation \cite{feynman1971lectures}. The central aspect is quantum complementarity --- the trade-off between quantum interference and which-way information. 
In the version discussed here, the setup is a protocol between Alice and Bob, whose systems interact gravitationally (see Figure \ref{fig:setup}). 
Alice performs an interference experiment with a massive particle which, in the distant past, she has prepared in a superposition of spatial separation $d$. Bob can choose to measure, or not to measure, Alice's gravitational field. Bob does that with a trapped massive particle that he can release or keep trapped. If Bob releases his particle, it will entangle to Alice's particle. Alice uses a time $T_A$ to recombine her interfering particle and measure its quantum interference. If Bob decides to open his trap, he measures the position of the released particle after a time $T_B$, giving which-way information on Alice's system, and thereby causing Alice's interference experiment to fail. Since opening the trap will affect Alice's outcome, but keeping the particle trapped will not, Bob can effectively send one bit of information to Alice. If the distance $D$ between Alice and Bob satisfies $cT_B<cT_A< D$, where $c$ is the speed of light, this would correspond to superluminal signalling. On the other hand, if Bob were able to measure his particle's position without decohering Alice's system, it would instead violate quantum complementarity: he would gain which-way information while Alice observes interference. Thus either quantum complementarity or no-signalling seems to be violated. Herein lies the apparent paradox. \\

\section{A resolution hinting at a quantum field theory of gravity}

A recent resolution \cite{Belenchia2018} involves two physical mechanisms,
which together imply the need to quantize the gravitational field. One mechanism is quantum fluctuations of the gravitational field, which limits Bob's ability to measure his particle's position to higher precision than the Planck length $l_p$. The other mechanism is quantized gravitational radiation, which will be emitted from Alice's interferometer and thus decohere the experiment if it is performed in a too short time period. In both cases, the relevant parameter is assumed to be the effective quadrupole moment $\mathcal{Q}_A \sim m d^2$ of Alice's superposed system. 
However, we will show below that the effect from the quadrupole vanishes in the considered setup, and the first relevant term is an octopole.
The effect is therefore typically weaker than assumed in \cite{Belenchia2018}. For the sake of presentation we will, however, proceed in this section as if the effect manifests itself at the quadrupole.
It quantifies emission of radiation on Alice's side, and is assumed in ref. \cite{Belenchia2018} to also quantify Bob's ability to distinguish Alice's two paths, since the difference in gravitational acceleration $\Delta g = g_R-g_L$ will not be larger than $\sim\frac{G \mathcal{Q}_A}{D^4}$, where $G$ is the gravitational constant.

Bob will get which-way information if his free-falling particle will differ in position by more than a Planck length for the two superposed paths of Alice,
\begin{equation}
    \Delta x > l_p. 
    \label{Bob_planck_length_limit}
\end{equation}
This gives a lower threshold for the strength of the effective quadrupole of Alice's particle:
\begin{equation}
    \frac{\mathcal{Q}_A}{m_p D^2} > \frac{D^2}{c^2T_B^2},
    \label{Bob_distinguishability}
\end{equation}
where $m_p = \sqrt{\frac{\hbar c}{G}}$ is the Planck mass. 
The right-hand side is larger than 1 since $cT_B<D$ in the paradox. 
Since the setup of the experiment implies $d<D$, we get the minimal lower bound when $cT_B \lesssim D$. This gives the threshold for Bob's ability to distinguish between Alice's two interferometric paths:
\begin{equation}
    \mathcal{Q}_A > m_p D^2.
    \label{eq:limitDistinguish}
\end{equation} 
For a quadrupole below this threshold, the difference in field strength from Alice's particle being on either side of the interferometer is not large enough for Bob to distinguish them. This prevents Bob from getting which-way information. \\ 

If, however, the quadrupole moment is large enough for eq. \eqref{eq:limitDistinguish} to be satisfied, the paradox remains unresolved. For this case, the authors of \cite{Belenchia2018} propose a solution which involves emission of radiation. As Alice's particle is accelerated for the two parts of the superposition to recombine, it will emit radiation which, if quantized, results in a number of gravitons
\begin{equation}
    N \sim \left(\frac{\mathcal{Q}_A}{m_pc^2T_A^2}\right)^2.
\end{equation}
By emitting radiation, Alice's system decoheres and her interference experiment fails, regardless of whether Bob opens his trap or not\footnote{For the electromagnetic case, decoherence due to Bremsstrahlung was carefully studied in ref. \cite{Breuer2001}}. To avoid decoherence we need $N < 1$, meaning there is also an upper limit on the quadrupole moment:
\begin{equation}
    \mathcal{Q}_A < m_pc^2T_A^2.
    \label{Alice_no_radiation}
\end{equation}
We thus have a lower limit on the mass quadrupole stemming from the required distinguishability on Bob's side, eq. \eqref{eq:limitDistinguish}, and an upper limit stemming from negligible emission of radiation on Alice's side, eq. \eqref{Alice_no_radiation}. 
Since the paradox arises for $cT_A<D$, 
both inequalities cannot be satisfied at the same time. Thus the two assumptions, namely inability to distinguish positions below the Planck length and the emission of gravitational radiation, resolve the paradox for all possible values of the quadrupole. 
\\

\section{A Resolution without quantization of gravity}

The above solution required the introduction of quantum fluctuations and quantum radiation of the gravitational field, strongly suggesting the need for a quantum field description of gravity \cite{Belenchia2018, belenchia2019information}. The Planck length enters this resolution as the size of quantum fluctuations in the gravitational field, giving rise to an uncertainty in the geodesic deviation equation. The Planck length can, however, also just be assumed as a lower cutoff beyond which no conclusions can be drawn, as was explicitly stated in \cite{Baym2009}. We now show that the latter assumption is entirely sufficient to resolve the paradox, without the need to consider radiation. Our resolution hence shows that the argument involving a quantum field theory of gravity is not necessary to resolve the paradox. 

Firstly, we analyse the specific setup and show that the effect is governed by the octopole.
In the multipole expansion of Alice's setup, with her particle moving one way and her lab moving ever so slightly the other way (see Appendix), the dipole moment vanishes as correctly pointed out in \cite{Belenchia2018}. However, the
quadrupole moment is equivalent for the two superposed cases that Bob aims to distinguish, whether Alice's particle is to the right or the left. This means that when Bob looks at the difference between these two cases to get which-way information, the first non-zero term contains the octopole moment $\mathcal{O}_A \sim md^3$. Therefore, the correct form of the condition for distinguishability, eq. \eqref{eq:limitDistinguish}, is actually
\begin{equation}
    \mathcal{O}_A > m_pD^3. 
    \label{eq:octopole_limit_distinguish}
\end{equation}
This bound is larger than the quadrupole bound by a factor of $D/d>1$. The generalization of the argument to higher multipoles has also been pointed out in footnote 11 of Ref. \cite{Belenchia2018}.

Secondly, building on Baym and Ozawa \cite{Baym2009}, we consider the measurements in Alice's interference experiment. Independent of the details of Alice's experiment, the fringe spacing $\delta_f$ in the interference pattern will generically depend on the wavelength $\lambda$ of the matter-wave, as well as the angle $\alpha$ at which the two beams are recombined, given by $\delta_f \sim \lambda/\alpha \sim \lambda\frac{L}{d}$. Thus the fringe spacing is 
\begin{equation}
    \delta_f \sim \lambda \frac{L}{d} = \frac{\hbar}{m_A v}\frac{vT_A}{d} = l_p\frac{m_p}{m_A}\frac{cT_A}{d},
    \label{fringe_spacing}
\end{equation}
where $L = v T_A$ is the characteristic length over which Alice's particle recoheres, at a speed $v$ over a time $T_A$. 
For the fringe pattern to be measurable, we need $\delta_f > l_p$, which gives
\begin{equation}
    \mathcal{O}_A < m_p d^2 cT_A.
    \label{Alice_fringes}
\end{equation}
This yields an upper bound on the octopole moment. Note that this bound is lower than the bound on the quadrupole found from the radiation argument, eq. \eqref{Alice_no_radiation}, by a factor $d/cT_A < 1$. 
Thus for masses for which radiation would decohere the interference, Alice could anyway not distinguish her interference pattern as it would be finer than the Planck length. 
In other words, whether or not Alice's particle radiates is irrelevant since she anyway cannot resolve her interference pattern. Consequently, the introduction of quantized radiation is redundant for solving the paradox. \\

One may ask if the above resolution that only relies on the Planck-length is generically valid, or if a modification of the Gedankenexperiment can restore the arguments in favor of quantized radiation and quantum fluctuations of the field as suggested in \cite{Belenchia2018}. A natural modification is for Bob to perform a better experiment that provides better which-way information of Alice's state. One possibility is to replace his free-falling particle with an interferometer, which measures the gravitational potential rather than the acceleration induced from Alice's particle. In this case, Bob is sensitive not to $\Delta g$, but to $\Delta V=V_R - V_L$, where $V_{R(L)}$ is the potential difference in the two sides of Bob's interferometer if Alice's particle is to the right (left). From this difference in potential, Bob can measure a relative phase shift $\phi_{R(L)}$ which is accumulated over the time of the experiment $T_B$. He will be able to get which-way information if he can distinguish between Alice's two cases, so he needs an accuracy greater than 
\begin{align}
    \Delta\phi = \phi_R-\phi_L \sim \frac{G\mathcal{O}_Am_BbT_B}{\hbar D^5} = \frac{\mathcal{O}_Am_BbcT_B}{m_p^2 D^5},
\end{align}
where $b$ is the spatial extent of his interferometer.

For different strengths of the potential from Alice's particle, the pattern of Bob's interference experiment will have the same fringe spacing $\delta_{f,B}$, but will be spatially shifted. Therefore, Bob can get which-way information my measuring the position of the peaks of his interference pattern, and from this position judging whether Alice's particle is to the right or left. The distance between where the fringe peaks would be for the two cases of Alice's particle is
\begin{align}
    \Delta &= \frac{\Delta\phi}{2\pi}\delta_{f,B} \sim  \frac{\mathcal{O}_Am_Bb}{m_p^2D^4} \frac{cT_B}{D}\lambda \frac{L_B}{b}\\ &=  \frac{\mathcal{O}_A}{m_p^2D^3} \frac{cT_B}{D}\frac{m_Bb}{m_pD}\frac{m_p}{m_B}\frac{cT_B}{b}l_p <  \frac{\mathcal{O}_A}{m_pD^3}
    l_p.
\end{align}
For Bob to get which-way information, this distance has to be larger than a Planck length $\Delta > l_p$, which gives a limit
\begin{align}
    \mathcal{O}_A > m_pD^3,
    \label{Bob_interfer_limit}
\end{align}
on the octopole moment. We see that this is the same limit as for the case of Bob having a free-falling particle found in eq. \eqref{eq:octopole_limit_distinguish}. In other words, eq. \eqref{Alice_fringes} and \eqref{Bob_interfer_limit} again resolve the paradox from the Gedankenexperiment, without the introduction of a quantum field theory of gravity. In this symmetric scenario it is clear that only a single assumption enters, namely the inability to resolve quantum interference finer than the Planck-length.
Whether this limitation is universally valid for any experimental setup is uncertain, but it resolves the Gedankenexperiment at hand. \\

\section{conclusions}
Our conclusion is that the question of the need to quantize gravity as a quantum field theory remains inconclusive, in contrast to the results found in \cite{Belenchia2018}. We also clarify an issue in previous treatments of this Gedankenexperiment, where it was assumed that the effect is governed by the effective mass dipole \cite{mari2016experiments, Baym2009} or quadrupole \cite{Belenchia2018, belenchia2019information}. In the most commonly considered scenario, however, the which-way information becomes relevant only on the octopole level, which places even more stringent constrains on the setups, as well as on experimental efforts that aim to probe gravitational entanglement between interfering particles in separate labs. In our resolution of the Gedankenexperiment
the only assumption used is the inability to resolve distances smaller than the Planck length, rather than referring to radiation. One may argue that this alone hints at a quantum theory with fluctuations on the Planck scale \cite{Padmanabhan1985}. However, without the need for quantized radiation the argument for a quantum field theory is weakened --- the cut-off also arises in other contexts \cite{hooft1993dimensional, Garay1995, Hossenfelder2013}. The Planck length cut-off could simply be taken as a limit on what we can predict with our current understanding of physics \cite{Baym2009} (in the words of A. D. Sakharov: ``It is natural to suppose also that [$l_p$] determines the limit of applicability of present-day notions of space and causality.'' \cite{sakharov1967vacuum}). Thus, the single assumption of a Planck length limit on measurability is not by itself a conclusive argument in favor of a quantum field theory of gravity. A Gedankenexperiment in clear favor of the necessity for a quantized gravitational field, in analogy to Bohr and Rosenfeld, remains lacking. The difficulty in formulating such an argument might suggest that our current understanding of physics is insufficient to address this question.
\\

\begin{acknowledgments}
We thank Alessio Belenchia, \v{C}aslav Brukner, and Gunnar Bj\"{o}rk for discussions and comments. This work was supported by the Swedish Research Council under grants no. 2019-05615 and 2020-04980. I.P. also acknowledges support by the European Research Council under Grant No. 742104 and The Branco Weiss Fellowship -- Society in Science.
\end{acknowledgments}

\bibliography{bibliography.bib}

\appendix*
\section{Which-way information obtained by Bob}
Bob's probe system, whether a falling particle or interferometer, will measure the gravitational field from Alice's side. This is governed by her gravitational potential $V(\vec{r}) = - G\int d^3r' \rho(r')/|\vec{r}-\vec{r'}|$, where $\rho(r')$ is the mass distribution of Alice's particle, her lab and herself.
Since all masses are on the same line, the multipole expansion of the potential becomes
\begin{equation}
\begin{split}
    V(r) &= -\frac{Gm_{tot}}{r} - \frac{G}{r^2} \mathcal{D}_A - \frac{G}{2r^3} \mathcal{Q}_A - \frac{G}{6r^4} \mathcal{O}_A - ...,
\end{split}
\end{equation}
where the origin is defined as the center of mass of Alice's system, and the total mass $m_{tot}$, the dipole moment $\mathcal{D}_A$, the quadrupole moment $\mathcal{Q}_A$, and the octopole moment $\mathcal{O}_A$ are properties of Alice's system.  

For the case of Alice's particle of mass $m_A$ being a distance $d$ to the right (left), her lab of mass $M_A$ will accordingly be moved a distance $\eta d $ to the left (right), where $\eta = \frac{m_A}{M_A}$.  The multipoles for the two cases are
\begin{align}
    m_{tot}&= m_A + M_A\\
    \mathcal{D}_A &= 0\\
    \mathcal{Q}_A & = 2m_A d^2 (1+ \eta)\\
    \mathcal{O}_A & =  \pm 6m_A d^3 (1 - \eta^2).
\end{align}   
The dipole moment disappears because of conservation of momentum, while the other moments are non-zero. However, the difference in the gravitational potential between the two cases only manifests itself at the octopole level.

Bob is trying to distinguish the two different gravitational fields. If he has an interferometer consisting of a particle of mass $m_B$ in a spatial superposition of separation $b$, he is directly sensitive to the difference in the two potential energies. The relative phase shift of his interferometer will be of order
\begin{equation}
 \phi \sim \frac{m_BT_B}{\hbar} \left(V(D)-V(D+b)\right),
\end{equation}
where $b \ll D$. The difference in phase shift, depending on the two superposed cases of Alice, is therefore 
\begin{equation}
 \Delta \phi \sim \frac{G m_B b T_B}{\hbar} \frac{\mathcal{O}_A}{D^5},
\end{equation} 
\\
which captures the distinguishability of Alice's superposition as measured by Bob.

\end{document}